# Quality-aware Approach for Engineering Self-adaptive Software Systems

Mohammed Abufouda

Department of Computer Science, Technical University of Kaiserslautern, Kaiserslautern, Germany
abufouda@cs.uni-kl.de

## ABSTRACT

Self-adaptivity allows software systems to autonomously adjust their behavior during run-time to reduce the cost complexities caused by manual maintenance. In this paper, an approach for building an external adaptation engine for self-adaptive software systems is proposed. In order to improve the quality of self-adaptive software systems, this research addresses two challenges in self-adaptive software systems. The first challenge is managing the complexity of the adaptation space efficiently and the second is handling the run-time uncertainty that hinders the adaptation process. This research utilizes Case-based Reasoning as an adaptation engine along with utility functions for realizing the managed system's requirements and handling uncertainty.

## KEYWORDS

*Software Quality, Model-Driven Software, Self-adaptive Software Systems, Case-based Reasoning,*

*Run-time Uncertainty*

## 1. INTRODUCTION

The majority of the existing work in the literature agrees [1] [2] that *self-adaptivity* in software systems is the ability of a software system to adjust its behaviour during run time to handle software system's complexity and maintenance costs [3] while preserving the requirement of the system. This property dictates the presence of an adaptation mechanism in order to build the logic of self-adaptivity without human intervention. Developing a self-adaptive software system is subjected to many challenges like handling the complexity of the adaptation space of the managed system. This complexity is conceived when the number of the states that the managed system can run in is relatively large. Also, this complexity manifests itself when new states are needed to be inferred from previous one i.e. learning from past experience. Another challenge is the uncertainty that hinders the adaption process during run-time. This paper will address these challenges. More precisely, our framework is concerned with the following problems:

- *Adaptation responsible unit:* The majority of the existing work do not provide a modular separation between the adaptation engine and the managed system. Embedding the adaptation logic within the managed system components increases the complexity in the development process of a self-adaptive software system. This also limits the reusability of the work achieved in one application to other applications or domains.

- *Run-time uncertainty handling*: Uncertainty is a challenge that exists not only in self-adaptive software systems but also in the entire software engineering field on different levels. Therefore managing uncertainty is an essential issue in constructing a self-adaptive software system as uncertainty hinders the adaptation process if it is not handled and diminished.

- *Adaptation space:* The adaptation process raises a performance challenge if the adaptation space is relatively large, particularly when new adaptations are required to be inferred. This challenge requires an efficient mechanism that guarantees learning new adaptations as well as providing the adaptation with satisfactory performance. This means that the adaptation engine's response should be provided as soon as an adaptation is issued since late adaptations provided by the adaptation engine could be futile.

The rest of this paper is structured as follows: Section 2 lists the related work and the existing gabs in the literature. Section 3 shows the expected contributions of our research and Section 4 describes our proposed solution and its model. Section 5 and Section 6 contains the progress and the future of our research, in particular the evaluation. This paper concludes in Section 7.

## 2. RELATED WORK

The body of literature in the area of self-adaptivity has provided a plethora of frameworks, approaches and techniques to enhance self-adaptivity that is widespread in many fields. This section contains the related work to our research. In the following sections, we will present the related work categorized according to the mechanisms used to support self-adaptivity.

### 2.1 Learning based adaptation

Salehie and Tahvildari [2] proposed a framework for realizing the deciding process performed by an external adaptation engine. They used knowledge base to capture the managed system's information namely domain information, goals and utility information. This is used in the decision-making algorithm, as they name it, which is responsible for providing the adaptation decision. In [5], Kim and Park provided a reinforcement learning-based approach for architecture-based self-managed software using both on-line and off-line learning. FUSION [6] was proposed by Elkhodary et al. to solve the problem of foreseeing the changes in environment, which hinders the adaptation during run time for feature-based systems using a machine learning technique. In [7], Mohamed-Hedi et al. provided a self-healing approach to enhance the reliability of web services. A simple experiment was used to validate their approach without empirical evidence.

### 2.2 Architecture & model based adaptation

RAINBOW [8] is a famous contribution in the area of self-adaptation based on architectural infrastructures reuse. RAINBOW monitors the managed system using abstract architectural models to detect any constraints violation. GRAF [9] was proposed for engineering self-adaptive software systems. The communication between the managed system and GRAF framework is carried out via interfaces. This approach has a performance overhead because GRAF reproduces a new adaptable version of the managed system. Similar to GRAF [9] Vogel and Giese [10] assumed that adaptation can be performed in two ways, parameter adaptation and structural adaptation. They provided three steps to resolve structural adaptation and used a self-healing web application as an example. Morin et al. [11] presented an architectural based approach for realizing software adaptivity using model-driven and aspect oriented techniques. The aim of this approach was to reduce the complexities of the system by providing architectural adaptation based solution. They provided model-oriented architectures and aspect models for feature designing and selection. Khakpour et al. [12] provided PobSAM, a model-based approach that is used to monitor, control and adapt the system behaviour using LTL to check the correctness of adaptation. Asadollahi et al. [13] presented StarMX framework for realizing self-management for Java-based applications. In their work they provided so called autonomic manager, which is an adaptation engine that encapsulates the adaptation logic. Adaptation logic was implemented by arbitrary policy-rule language. StarMX uses JMX and policy engines to enable self-management. Policies were used to represent the adaptation behaviour. This framework is restricted to Java-based application as the definition of processes is carried out by implementing certain Java interfaces in the policy manager. They evaluated their framework against some quality attribute. However, their evaluation for quality attributes was not quantified quantitatively. The work in [14] provided a new formal language for representing self-adaptivity for architecture-based self-adaptation. This language was used as an extension of the RAINBOW framework [8]. This work explains the use of this new language using an adaptation selection example that incorporate some stakeholders' interests in the selection process of the provided service which represents the adaptive service. Bontchev et al. [15] provides a software engine for adaptable process controlling and adaptable web-based delivered content. Their work reuses the functionality of the existing component in order to realize self-adaptivity in architecture-based systems. This work contains only the proposed solution and the implementation without experiment and evaluation.

### 2.3 Middleware based adaptation

In [16], a prototype for seat adaptation was provided. This prototype uses a middleware to support an adaptive behaviour. This approach was restricted to the seat adaptation which is controlled by a software system. Adapta framework [17] was presented as a middleware that enabled self-adaptivity for components in distributed applications. The monitoring service in Adapta monitored both hardware and software changes.

### 2.4 Fuzzy control based adaptation

Yang et al. [18] proposed a fuzzy-based self-adaptive software framework. The framework has three layers: (1) Adaptation logic layer, (2) Adaptable system layer, which is the managed system and (3) Software Bus. The adaptation logic layer represents the adaptation engine that includes the fuzzy rule-base, fuzzification and de-fuzzification components. This framework has a set of design steps in order to implement the adaptation. POISED [19] introduced a probabilistic approach for handling uncertainty in self-adaptive software systems by providing positive and negative impacts of uncertainty. An evaluation experiment had been applied which showed that POISED provided an accepted adaptation decision under uncertainty. The limitations of this approach are that it handles only internal uncertainty and does not memorize and utilize previous adaptation decisions.

### 2.5 Programming framework based adaptation

Narebdra et al. [20] proposed programming model and run time architecture for implementing adaptive service oriented. It was done via a middleware that solves the problem of static binding of services. The adaptation space in this work is limited to three situations that require adaptation of services. MOSES approach was proposed in the work [21] to provide self-adaptivity for SOA systems. The authors used linear programming problem for formulating and solving the adaptivity problem as a model-based framework. MOSES aimed to improve the QoS for SOA, and the work in [21] provides a numerical experiment to test their approach. QoSMOS [22] provided a tool-supported framework to improve the QoS for the service based systems in adaptive and predictive manner. The work in [23] provided an implementation of architecture-based self-adaptive software using aspect oriented programming. They used a web-based system as an experiment to test their implementation. Their experiment showed that the response time of the self-adaptive implementation is better than the original implementation without a self-adaptivity mechanism. Liu and Parashar [24] provided Accord, which is a programming framework that facilitates realizing self-adaptivity in self-managed applications. The usage of this framework was illustrated using forest fire management application.

Table 1, which is similar to what proposed in [4], summarizes the related work done in this research. The table has two aspects of comparison (1) Research aspects and (2) Self-adaptivity aspect. The earlier aspect is important and represent an indication regarding the maturity and creditability of the research. The later aspect is related to the topic of this paper.

| Covered literature categorization | Work | Research aspects | | | | | | Self-adaptive software system aspects | | | | | |
|---|---|---|---|---|---|---|---|---|---|---|---|---|---|
| | | Problem Statement | Contribution statement | Experiment | evaluation metrics | Limitations | Threats to validity | Adaptation Expediency | Adaptation remembrance | Uncertainty Handling | Adaptation Res. Time | Adaptation style | Adaptation engine |
| Learning based adaptation | [2] | √ | √ | X | X | X | X | X | √ | X | X | Dynamic | External |
| | [5] | √ | √ | √ | X | X | X | √ | X | X | X | Dynamic | External |
| | [6] | √ | √ | √ | √ | √ | X | √ | √ | X | √ | Dynamic | External |
| | [7] | X | X | √ | X | X | X | X | X | X | X | Dynamic | External |
| Architecture & model based adaptation | [8] | √ | √ | √ | √ | X | X | X | X | √ | √ | Dynamic | External |
| | [9] | √ | √ | √ | √ | X | √ | X | X | X | X | Dynamic | External |
| | [10] | √ | √ | √ | X | X | X | X | X | X | X | Static | Internal |
| | [13] | X | X | √ | X | X | X | √ | X | X | X | Dynamic | External |
| | [11] | X | X | √ | √ | X | X | √ | X | X | √ | Dynamic | External |
| | [12] | √ | √ | X | X | X | X | X | X | X | X | Dynamic | Internal |
| | [14] | √ | √ | √ | X | X | X | X | X | X | X | Static | External |
| | [15] | √ | √ | X | X | X | X | X | √ | X | X | Dynamic | External |
| Middleware based adaptation | [16] | √ | √ | √ | X | X | X | √ | X | X | X | Static | Internal |
| | [17] | √ | √ | X | X | X | X | X | X | X | X | Dynamic | External |
| Fuzzy control based adaptation | [18] | √ | √ | X | X | X | X | X | X | X | X | Dynamic | External |
| | [19] | √ | √ | √ | √ | X | X | √ | X | √ | √ | Dynamic | Internal |
| Programming framework based adaptation | [20] | X | X | √ | √ | X | X | X | X | X | X | Dynamic | External |
| | [21] | √ | √ | √ | X | X | X | √ | X | X | X | Dynamic | External |
| | [23] | √ | √ | √ | √ | X | X | √ | X | X | √ | Dynamic | Internal |
| | [24] | √ | √ | √ | X | X | X | √ | X | X | √ | Dynamic | Internal |

Table 1: Summary of related work

## 3. RESEARCH CONTRIBUTION

In this research, we realize self-adaptivity in software system by providing an external adaptation engine which reduces the changes in the managed system and subsequently in the entire self-adaptive system. Our approach utilizes Case-based Reasoning (CBR) [25] as an external adaptation engine in order to overcome the aforementioned challenges. Specifically, this research proposes a framework that we claim it addresses the following challenges:

- Separating the managed system and the adaptation engine in a modular fashion in order to overcome the drawbacks of embedding the self-adaptivity logic within the managed system. This idea is one of the key ideas in the IBM autonomic element [26] which suggests a modular separation between the managed system and the adaptation engine.

- Managing the complexity of adaptation space by remembering the previously achieved adaptations stored in a knowledge base, which improves the performance of the adaptation process. The remembrance supports not only the complexity of the adaptation space, but also the performance of the adaptation engine. That is because recalling already existing adaption is better than constructing it from scratch in terms of performance.

- Handling the run-time uncertainty that appears in the adaptation process due to the managed system's environment changes or our framework's internal model. We utilize and incorporate the probability theory and the utility functions as proposed in [27].

## 4. PROPOSED SOLUTION

In this section, an overview of our proposed solution will be presented. Figure 1 shows a reference model of our solution that will be described in the following sections.

### 4.1. External adaptation engine

The adaptation engine contains an *adaptation mediator*, which is responsible for:

- Monitoring the managed system by reading its attributes to decide whether an adaptation is required or not. We suppose that the managed system provides a service with overall utility $U$. If $U$ is below or is approaching a predefined utility threshold i.e. "$UT$", then the monitoring unit issues an adaptation process. The *adaptation request* is the set of the managed system attribute values at the time of issuing the adaptation. Consequently, the adaptation request is sent to the adaptation engine to perform the adaptation process.

- Executing the *adaptation response* received by the adaptation engine. The adaptation response is the result of the adaptation process performed by the adaptation engine, which is the corrective state to be applied on the managed system.

In addition to the adaptation mediator, the adaptation engine embraces a *Case-based reasoning engine*. Typically, CBR life cycle consists of four stages:

1. *Retrieve*: The CBR system retrieves the most similar case(s) from the Knowledge Base (KB) by applying the similarity measures on the request case.

2. *Reuse (Adapt):* In this stage, CBR employs the information of the retrieved cases. If the retrieved cases are not sufficient in themselves to solve the query case, the CBR engine adapts this/these case/s to generate a new solution. Some of the common techniques for reusing and adapting the retrieved knowledge are introduced in [28]. Our approach uses *Generative Adaptation* [29], which requires some heuristics e.g. utility functions, to provide an efficient adaptation process.

3. *Revise:* A revision of the new solution is important to make sure that it satisfies the goals of the managed system. Revision process can be done by applying the adaptation response to real world, evaluate it by the domain expert or by simulation approaches.

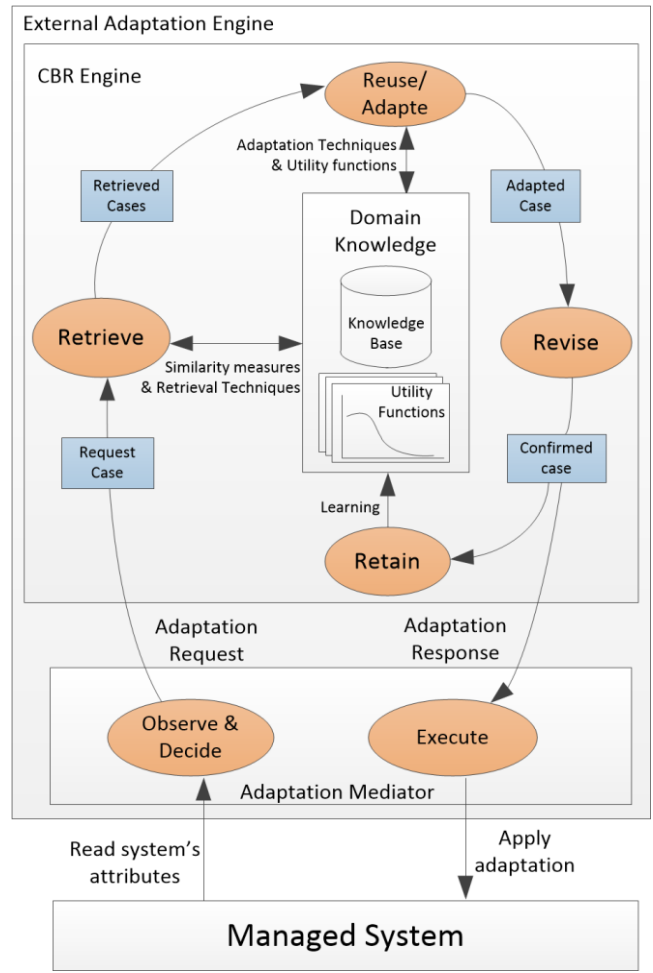

Figure 1: Reference modle for the proposed solution

4. *Retain*: In this stage, the new generated cases are saved in the knowledge base. Case-Based Learning (CBL) have been introduced in [30] to provide algorithms and approaches for the retain process.

In our model, the case is a set of attributes that represents the attributes of the managed system. For example, if one attribute of the managed system causes a *UT* break, then the adaptation engine should alter the value of this attribute in order to provide the required utility. In our solution we incorporate the utility functions for capturing the requirements of the managed system. Also, utility function is used to judge the quality of the cases stored in the KB and generated by the adaptation engine.

Algorithm 1 abstracts the adaptation process of our solution, where β is a predefined level of the accepted similarity and QAF is the qualified adaptation frame, a set of cases that have the potential to be used directly as an adaptation response or as basis for adaptation. *Case Expediency* is a measure for the usefulness of a case, and this measure uses the similarity of the case beside its utility.

---

**Algorithm 1** Adaptation process

**Require:** $KB$, $A_{req}$
**Ensure:** $Utility(A_{res}) > UT$
1: List cases $\Leftarrow$ Retrieve ($KB, A_{req}$)
2: List QAF
3: Case $A_{res}$
4: **while** Case $c \Leftarrow$ Iterate(cases) **do**
5:    **if** $Sim(A_{req}, c) \in [\beta, 1]$ **then**
6:       QAF.add(c)
7:    **end if**
8: **end while**
9: **if** QAF is not *Empty* **then**
10:    $A_{res} \Leftarrow max(CaseExpediency(QAF))$
11:    **Return** $A_{res}$
12: **else**
13:    $A_{res} \Leftarrow ConstructiveAdapt(A_{req})$
14:    Retain($A_{res}, KB$)
15:    **Return** $A_{res}$
16: **end if**

---

Algorithm 1: Adaptation process algorithm

### 4.2. Uncertainty quantification

We follow the uncertainty quantification approach in [31], where uncertainty has three dimensions:

- The Location of uncertainty: Where the uncertainty manifests in the system.

- The Level of uncertainty: A variation between deterministic level and total ignorance. This means that uncertainty about one attribute of the system can take a value between one and zero.

- The Nature of uncertainty: Whether the cause of uncertainty is variability or lack of knowledge in the uncertainty meant attribute of the system.

Based on [32], uncertainty in self-adaptive software systems can be found in three places, namely: System requirement, system design and architecture, and run-time. In our solution, we focused on run-time uncertainty by quantifying it based on the aforesaid three dimensions.

## 5. PROGRESS AND CURRENT STATUS

A prototypical implementation of the solution has been done. This implementation includes the integration of the CBR engine with utility functions. The implementation also includes the generative adaptation of the adaptation responses. Moreover, uncertainty analysis and quantification are provided in this implementation paving the way for handling uncertainty during run-time. The three dimensions of the uncertainty [31] has been modelled and implemented.

## 6. FUTURE DIRECTION AND EVALUATION

For future direction, firstly, we will use a case study to empirically evaluate and validate our approach. The case study i.e. the managed system, should require the self-adaptivity mechanism that performs well under run-time uncertainty. Secondly, we will evaluate the results of the case study application. The evaluation will be based on software quality metrics and GQM [33]. We expect that the experimentation of our solution will provide a positive potential results for both handling the uncertainty and the complexity of adaptation space. However, we do not have a clue regarding the response time of the adaptation engine, the results will reveal this issue.

## 7. CONCLUSION

In this paper, we have presented our research for realizing self-adaptivity in software systems. We started by showing the gabs in the research and the expected contributions of the research. Also, we have presented details about the solution model and the used technology, Case-based reasoning. The progress of the work was presented along with the future directions. This paper ended with our vision of the evaluation process of the solution.

**Authors**

**Mohammed Abufouda** received the BSc. in Computer Engineering from Islamic University, Palestine in 2006 and the MSc. degree in computer science from Technical University of Kaiserslautern, Germany, in 2013. He is a PhD candidate at Technical University of Kaiserslautern in computer science department. His research interests includes software engineering and complex system analysis.

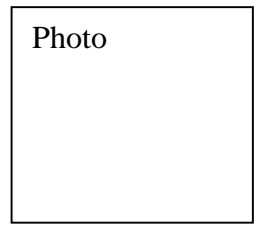

Photo